\documentclass[12pt]{iopart}

\usepackage{iopams} 

\begin{document}

\bibliographystyle{unsrt}

\title[]{Raising and lowering operators, factorization and differential/difference operators
of hypergeometric type}

\author{Miguel Lorente}

\address{Departamento de F\'{\i}sica, Universidad de Oviedo, 33007 Oviedo, Spain}

\begin{abstract}

Starting from Rodrigues formula we present a general construction of raising and
lowering operators for orthogonal polynomials of continuous and discrete variable on uniform lattice. In order
to have these operators mutually adjoint we introduce orthonormal functions with respect
to the scalar product of unit weight. Using the Infeld-Hull factorization method, we
generate from the raising and lowering operators the second order self-adjoint
differential/difference operator of hypergeometric type.
\end{abstract}

\noindent PACS: 02.10Nj; 02.20.Sv; 02.70.Bf; 03.65.Ge

\section{Introduction}
The factorization method has become a very powerful tool to solve second order
differential equations and its application to physical models with orthonormal basis,
generated by creation and annihilation operators. A classical paper by Infeld and Hull [1]
defined the method and applied it to a large class of second order Hamiltonian that
generalizes the well known description of the non-relativistic oscillator by means of
creation and annihilation operators. Miller [2] enlarged this method to difference
equation and made connection to orthogonal polynomial of discrete variable. An analysis of
the factorization types led Miller to the idea that this method is a particular case of
the representation theory of Lie algebras.

The two volumes of Nikiforov,  and colaborators [3] [4] about classical orthogonal polynomials of
continuous and discrete variable opened the way to a more rigurous and systematic approach
to the factorization method. In fact Atakishiev and colaborators [5] [6] [7] [8] explored
the application of Kravchuk, Meixner and Charlier polynomials to the eigenvalue problem of
some dynamical system where the energy eigenvalues are equally spaced. This particular situation makes possible to determine the
generators of the dynamical symmetry group. Also Smirnov [9] has used the properties of difference equations of
hypergeometric type given in [3] to construct raising and lowering operators that generate
orthonormal functions corresponding to Hamiltonian of different levels.

Bangerezako and Magnus have developed the method of the factorization of difference
operators of hypergeometric type [10] [11] [12]. They proposed two different approaches
for this factorization. 1) For a given operator find raising and lowering operators that generate a
complete set of polynomials eigenfunctions. 2) Generate from a factorization chain an operator having a
complete set of polynomial eigenfunctions.

We have presented two papers [13] [14] related to the construction of the creation and
annihilation operators for the orthogonal polynomials (or functions) of continuous and
discrete variables. The motivation for these papers was the construction of mathematical
model for Quantum systems on discrete space-time (as the harmonic oscillator, the hydrogen
atom, the Dirac equation) [15] and to make connection with standard quantum mechanics by the
continuous limit.

In this paper we follow the second approach to the factorization method of Bangerezako and
Magnus explained before. Starting from the raising and lowering operators we generate the
second order differential/difference equation corresponding to the hypergeometric
functions of continuous and discrete variable. Our procedure is completely general and
valid for all functions of this type.

In section 2 we use the results of Nikiforov, Suslov and Uvarov [3] [4] connecting first
order derivatives and orthogonal polynomials (as a consequence of Rodrigues' formula) to
construct raising and lowering operators (the last one with the help of recurrence
relations). In general these operators are not mutually adjoint with respect to the
standard scalar product. For this reason, we introduce in section 3 the orthonormalized
functions of hypergeometric type and then the corresponding raising and lowering operators
are always mutually adjoint.

In section 4 and 5 we repeat the same systematic procedure, derived from Rodrigues's
formula, to calculate the raising and lowering operators for orthogonal polynomials and
functions of discrete variable. It can be proved that these operators are also mutually
adjoint.

In section 6 we introduce the factorization method to generate the second order
differential operator of the Sturm-Liouville type having a complete set of polynomials
eigenfunctions. The factorization of the raising and lowering operators fulfills (up to
a factor) the defining equations of the Infeld-Hull method [1].

In section 7 we apply the same technique to the hypergeometric functions of discrete
variable. As usual all these functions transform in the limit into the corresponding hypergeometric functions of continuous
variable.

It is important to make clear that the raising and lowering operators, introduced in sections 2 to 5,
are defined with respect to one index only, namely, the degree of the orthogonal polynomials or the
degree of the corresponding orthonormal functions. The same definition has been used by Atakishiyev and
colaborators [6], [7], [8], by Bangerezako and Magnus [10], [11], [12] and by Infeld, Hull and Miller
[1], [2].
Physically this situation corresponds in the case of quantum oscillators to the creation and
annihilation operators with respect to the index that distingues different eingenvectors of the energy
operators.

With respect to the factorization techniques in the case of difference equations of section 7 two types
of factorization can be considered [12]. Writting a linear difference equation of second order in the
form

\[H(x)y(x)=\sum\limits_{i=-d}^d {A_i}(x)E_x^iy(x)\]
where $E_x^i\left[ {f(x)} \right]=f(x+i),d\in Z^+,i\in Z$
  and ${A_i}(x)$ are some scalar functions in $x$, the first type of factorization consists in
factorizing exactly the operators $H(x)+C$, with $C$ some constant, and the raising and lowering
operators satisfying a quasi-periodicity condition (Spiridinov-Vinet-Zhedanov type) [19]. The second
factorization technique consists in factorizing the operator$E_x^d\circ \left[ {H(x)+C} \right]$
 with some raising and lowering operators that are shape-invariant (Infeld-Hull-Miller type) [1], [2]. In
section 7 we have used the first type of factorization, but in section 6 obviously we have used the
Infeld-Hull-Miller technique for differential equations of hypergeometric type.

\section{Raising  and lowering operators for orthogonal polynomials of continuous
\mbox{variable}}

A polynomial of hypergeometric type $y_n (s)$
 of continuous variable $s$
 satisfies two fundamental equations; from which one derives raising and lowering operator

\medskip
\noindent i) Differential equation
\[\fl ({\rm C1})\qquad  \sigma (s)y''_n  + \tau (s)y'_n (s) + \lambda _n y_n (s) = 0\]
where $\sigma (s)$ and $\tau (s)$ are polynomials of, at most, second and first degree
respectively, and $\lambda _n $ is a constant, related to the above functions
\[\lambda _n  =  - n\left( {\tau ' + \frac{{n - 1}}{2}\sigma ''} \right)\]

The differential equation can be written in the form of an eigenvalue equation of
Sturm-Liouville type:
\[\left( {\sigma (s)\rho (s)y'_n (s)} \right)^\prime   + \lambda _n \rho (s)y_n (s) = 0\]
where $\rho (s)$ id the weight function, satisfying
$\left( {\sigma (s)\rho (s)} \right)^\prime   = \tau (s)\rho (s)$ .

The solutions of the differential equation are polynomials that satisfy an orthogonality
relation with respect to the scalar product
\[\int_a^b {y_n } (s)y_m (s)\rho (s)ds = d_n^2 \delta _{nm}\]
where $d_n $ is a normalization constant.

The differential equation (C1) defines an operator that is self adjoint with respect to
this scalar product

\medskip
\noindent ii) Three term recurrence relations:
\[\fl ({\rm C2)}\qquad sy_n (s) = \alpha _n y_{n + 1} (s) + \beta _n y_n (s) + \gamma _n y_{n
- 1} (s)\]
where $\alpha _n ,\beta _n ,\gamma _n $ are constants.

\medskip
\noindent iii) Raising operator: From the Rodrigues formula (which is a consequence of the
differential equation (C1) one derives a relation for the first derivative of polynomials
$y_n (s)$ in terms of the polynomials themselves
\[\sigma y'_n (s) = \frac{{\lambda _n }}{{n\tau '_n }}\left[ {\tau _n (s)y_n (s) -
\frac{{B_n }}{{B_{n + 1} }}y_{n + 1} (s)} \right]\]
where 
\begin{eqnarray*}
\tau _n (s) &=& \tau (s) + n\sigma '(s) \\
\tau '_n (s) &=& \tau ' + n\sigma '' =  - \frac{{\lambda _{2n + 1} }}{{2n + 1}}
\end{eqnarray*}

We can modify the last equation in a more suitable form. From
\[a_n  = B_n \prod\limits_{k = 0}^{n - 1} {\left( {\tau ' + \frac{1}{2}(n + k - 1)\sigma
''} \right)} \quad ,\quad a_0  = B_0 \]
we can prove the following identity
\[\alpha _n  = \frac{{a_n }}{{a_{n + 1} }} = \frac{{B_n }}{{B_{n + 1} }}\frac{{\tau ' +
\frac{{n - 1}}{2}\sigma ''}}{{\left( {\tau ' + \frac{{2n - 1}}{2}\sigma ''} \right)(\tau '
+ n\sigma '')}} = \frac{{B_n }}{{B_{n + 1} }}\frac{{\lambda _n }}{n}\frac{{2n}}{{\lambda
_{2n} }}\frac{{2n + 1}}{{\lambda _{2n + 1} }}\]
from which we finally get
\[
\fl ({\rm C3)}\qquad  + \frac{{\lambda _n }}{n}\frac{{\tau _n (s)}}{{\tau '_n }}y_n (s) -
\sigma (s)y'_n (s) = \frac{{\lambda _{2n} }}{{2n}}\alpha _n y_{n + 1} (s)
\]

The left side of this equation can be considered the differential operator which, when
applied to $y_n (s)$ , gives a polynomial of higher degree.

\medskip
\noindent iv) Lowering operator:

Introducing (C2) in (C3) we get a differential operator which, when applied to an
orthogonal polynomial of some degree, gives another polynomial of lower degree.

\[
\fl ({\rm C4)}\qquad  - \frac{{\lambda _n }}{n}\frac{{\tau _n (s)}}{{\tau '_n }} +
\frac{{\lambda _{2n} }}{{2n}}(s - \beta _n ) + \sigma (s)y'_n (s) = \frac{{\lambda _{2n}
}}{{2n}}\gamma _n y_{n - 1} (s)
\]

Formula (C3) and (C4) can be used to calculate solutions of the diferential equation (C1).
In fact, if we put $n = 0$ in (C4) we get $y_0 (s)$ . Inserting this value in (C3) we
obtain by iteration all the solutions of the differential operator (C1).

The explicit expressions for orthogonal polynomials of continuous variable are given in
Table I. The values of $\rho (s), \sigma (s), \tau (s), \lambda _n, \alpha _n, \beta _n, \gamma
_n, d_n$ are taken from [4].

\section{Raising and lowering operators for orthonormal functions of continuous variable}

From the orthogonal polynomials that satisfy a scalar product with respect to the weight
$\rho (s)$ we can construct a new functions:
\[\psi _n (s) \equiv d_n^{ - 1} \sqrt {\rho (s)} y_n (s)\]
and obtain orthogonal functions of unit norm. Solving the last expression for $y_n (s)$
 and substituing in (C1), (C2), (C3) and (C4) and using the properties of $\sigma (s)$
 and $\tau (s)$
 we obtain the following expressions for the normalized orthogonal functions:

\medskip
\noindent i) Differential equation:
\[ \fl ({\rm NC}1)\quad \sigma (s)\psi ''_n (s) + \sigma '(s)\psi '_n (s) - \left[
{\frac{1}{4}\frac{{\left( {\tau (s) - \sigma '(s)} \right)^2 }}{{\sigma (s)}} +
\frac{1}{2}\left( {\tau ' - \sigma ''} \right)}\right]\psi _n (s) + \lambda _n
\psi _n (s) = 0
\]
which corresponds to a self-adjoint operator of Sturm-Liouville type.

\medskip
\noindent ii)  Recurrence relation:
\[
 \fl({\rm NC2)}\quad \frac{{\lambda _{2n} }}{{2n}}\frac{{d_{n + 1} }}{{d_n }}\alpha _n \psi
_{n + 1} (s) + \frac{{\lambda _{2n} }}{{2n}}\frac{{d_{n - 1} }}{{d_n }}\gamma _n \psi _n
(s) + \frac{{\lambda _{2n} }}{{2n}}(\beta _n  - s)\psi _n (s) = 0
\]
\medskip
\noindent iii)  Raising and lowering operators:

\begin{eqnarray*}
\fl({\rm NC3)}\quad L^ +  (s,n)\psi _n (s) = \left[ {\frac{{\lambda _n }}{n}\frac{{\tau _n
(s)}}{{\tau '_n }} + \frac{1}{2}\left( {\tau (s) - \sigma '(s)} \right)} \right]\psi _n
(s) - \sigma (s)\psi '_n (s) \\
\lo = \frac{{\lambda _{2n} }}{{2n}}\alpha _n \frac{{d_{n + 1}
}}{{d_n }}\psi _{n + 1} (s) \\
\fl ({\rm NC4)}\quad L^ -  (s,n)\psi _n (s) = \left[ - {\frac{{\lambda _n
}}{n}\frac{{\tau _n (s)}}{{\tau '_n }}  + \frac{{\lambda _{2n} }}{{2n}}(s -
\beta _n ) - \frac{1}{2}\left( {\tau (s) - \sigma '(s)} \right)} \right]\psi _n (s)\\
\lo   +\sigma (s)\psi '_n (s) = \frac{{\lambda _{2n} }}{{2n}}\gamma _n \frac{{d_{n - 1}
}}{{d_n }}\psi _{n - 1} (s)
\end{eqnarray*}

Putting $n = 0$ in (NC4) we obtain $\psi _0 (s)$ , and inserting this value in (NC3) we
obtain by iteration all the orthonormal function of hypergeometric type.

The explicit expressions for these functions are given in Table I.

We want to make two observations. First, the operator corresponding to (NC1) is a
self-adjoint operator of Sturm-Liouville type as can be easily checked. Secondly, the
raising and lowering operators (NC3) and (NC4) are mutually adjoint in the case of
Laguerre and Hermite functions. For the Jacobi and Legendre functions we have to multiply both operators by $2n/\lambda_{2n}$.
In fact, we have
\begin{eqnarray*}
 \int\limits_a^b {\psi _{n + 1} (s)\left[ {\frac{{2n}}{{\lambda _{2n} }}L^ + 
(s,n)\psi _n (s)} \right]} \,ds = \alpha _n \frac{{d_{n + 1} }}{{d_n }}\\
\int\limits_a^b {\left[ {\frac{{2n + 2}}{{\lambda _{2n + 2} }}L^ -  (s,n + 1)\psi
_{n + 1} (s)} \right]} \psi _n (s)ds = \gamma _{n + 1} \frac{{d_n }}{{d_{n + 1} }}
\end{eqnarray*}

Both integrals are equal because $\gamma _{n + 1}  = \alpha _n \frac{{d_{n + 1}^2
}}{{d_n^2 }}$

\noindent (In the case of Hermite and Laguerre functions $\lambda _m/m$
is independent of $m$, for any $m$)

\section{Raising and lowering operators for orthogonal polynomials of discrete variable}

A polynomial of hypergeometric type $P_n (x)$ of discrete variable $x$ satisfies two
fundamental relations from which one derives raising and lowering operators.

\medskip
\noindent i)  Difference equation:
\[
\fl{({\rm D}1)} \quad  \sigma (x)\Delta \nabla P_n (x) + \tau (x)\Delta P_n (x) + \lambda _n P_n
(x) = 0
\]
where $\sigma (x)$ and $\tau (x)$ are polynomials of, at most, second and first degree,
respectively. The forward (backward) difference operators are:
\[\Delta f(x) = f(x + 1) - f(x)\quad \nabla f(x) = f(x) - f(x - 1)\]

This difference equation can be written in the form of an eigenvalue equation of Sturm
Liouville type
\[\Delta \left[ {\sigma (x)\rho (x)\nabla P_n (x)} \right] + \lambda _n \rho (x)P_n (x) =
0\]
where $\rho (x)$ is a weight function satisfying
\[\Delta \left[ {\sigma (x)\rho (x)} \right] = \tau (x)\rho (x)\] 
and $\lambda _n $ is the eigenvalue corresponding to the eigenfunction $P(x)$ :
\[\lambda _n  =  - n\Delta \tau (x) - \frac{{n(n - 1)}}{2}\Delta ^2 \sigma (x) =  - n\left(
{\tau ' + \frac{{n - 1}}{2}\sigma ''} \right)\]

The solution of the difference equation are polynomials that satisfy an orthogonality
relation with respect to the scalar product
\[\sum\limits_{x = a}^{b - 1} {P_n (x)} P_m (x)\rho (x) = d_n^2 \delta _{nm}\]
when $\delta _{mn}$ is the Kronecker symbol and $d_n$ some normalization constant. The
difference equation (D1) defines an operator that is self-adjoint with respect to this
scalar product.

\medskip
\noindent ii)   Three term recurrence relations:
\[
\fl ({\rm D2)}\quad xP_n (x) = \alpha _n P_{n + 1} (x) + \beta _n P_n (x) + \gamma _n P_{n
+ 1} (x)
\]
where $\alpha _n ,\beta _n ,\gamma _n $ are some constants.

\medskip
\noindent iii)    Raising operator
\smallskip

From the Rodrigues formula, one derives a relation for the first difference operator of
polynomials $ P_n (x)$ in terms of the polynomials themselves.
\[\sigma (x)\nabla P_n (x) = \frac{{\lambda _n }}{{n\tau '_n }}\left[ {\tau _n (x)P_n (x) -
\frac{{B _n }}{{B _{n + 1} }}P_{n + 1} (x)} \right]\]
where
\begin{eqnarray*}
 \tau _n (x) &=& \tau (x + n) + \sigma (x + n) - \sigma (x) \\ 
  \Delta \tau _n (x) &=& \Delta \tau (x) + n\Delta ^2 \sigma (x) \\
  {\rm or}\;\;\;\tau '_n  &=& \tau ' + n\sigma ''(x) =  - \frac{{\lambda _{2n + 1} }}{{2n +
1}}
\end{eqnarray*}
	because $\sigma (x)$ and $\tau (x)$ are polynomials of at most second and first order
degree respectively.

We can modify the last equation to a more suitable form, as we did in the continuous case.
From the definition
\[a_n  = B_n \prod\limits_{k = 0}^{n - 1} {\left( {\tau ' + \frac{1}{2}(n + k - 1)\sigma
''} \right)} ,\quad a_0  = B_0 \]
we have the following identity:
\[\alpha _n  = \frac{{a_n }}{{a_{n + 1} }} =  - \frac{{2n}}{{\lambda _{2n} }}\frac{{(2n +
1)}}{{\lambda _{2n + 1} }}\frac{n}{{\lambda _n }}\frac{{B_n }}{{B_{n + 1} }}\]
from which we get a more simplified version

\begin{equation*}
({\rm D}3)\quad \sigma (x)\nabla P_n (x) = \frac{{\lambda _n }}{n}\frac{{\tau _n
(x)}}{{\tau '_n }}P_n (x) - \frac{{\lambda _{2n} }}{{2n}}P_{n + 1} (x)
\end{equation*}

This equation defines the raising operator in terms of the backward difference.

\medskip
\noindent iv) Lowering operator
\smallskip

From the expresion for the raising operator we can derive another lowering operator in
terms of the forward operator. We substitute the difference operator $\nabla $
 in (D3) for its equivalent $\nabla  = \Delta  - \nabla \Delta $ , and then the difference
equation (D1) and the three terms recurrence relations (D2), with the result
\begin{eqnarray*}
\fl ({\rm D}4)\quad \left( {\sigma (x) + \tau (x)} \right)\Delta P_n (x) = \left[ { -
\frac{{\lambda _n }}{n}\frac{{2n + 1}}{{x_{2n + 1} }}\tau (x) - \lambda _n  -
\frac{{\lambda _{2n} }}{{2n}}(x - \beta _n )} \right]P_n (x) \\
+ \frac{{\lambda _{2n}
}}{{2n}}\gamma _n P_{n - 1} (x)
\end{eqnarray*}

As in the continuous cas from (D4) putting $n = 0$ we get $P_0 (x)$ and inserting this
value in (D3) we obtain by iteration all the polynomials $P_n (x)$ satisfying (D1).

The explicit expressions for the orthogonal polynomials $P_n (x)$ are given in table II.
The values of $\rho (x), \sigma (x), \tau (x), \lambda _n, \alpha _n, \beta _n, \gamma
_n, d_n$ are taken from [4].

\section{Raising and lowering operators for orthonormal functions of discrete variable}

In the last section we have a set of polynomials that are orthogonal with respect to the
weight function $\rho (x)$ . From these polynomials we construct some functions that are
orthogonal with respect to the unit weight, $\rho (x) = 1$ , and at the same time are
normalized.
\[\phi _n (x) = d_n^{ - 1} \sqrt {\rho (x)} P_n (x)\]

Introducing this expresion in (D1), (D2), (D3) and (D4) and using the properties of
function $\sigma (x)$, $\tau (x)$ and $\rho (x)$, we obtain

\medskip
\noindent i)  Difference equation
\begin{eqnarray*}
\fl ({\rm ND}1)\quad  \sqrt {\left( {\sigma (x) + \tau (x)} \right)\sigma (x + 1)} \phi _n
(x + 1) +\sqrt {\left( {\sigma (x - 1) + \tau (x - 1)} \right)\sigma (x)} \phi _n (x -
1) \\
- \left( {2\sigma (x) + \tau (x)} \right)\phi _n (x) + \lambda _n \phi _n (x) = 0
\end{eqnarray*}

\noindent ii)  Three term recursion relation:
\begin{eqnarray*}
\fl ({\rm ND}2)\quad \frac{{\lambda _{2n} }}{{2n}}\alpha _n \frac{{d_{n + 1} }}{{d_n }}\phi
_{n + 1} (x) + \frac{{\lambda _{2n} }}{{2n}}\gamma _n \frac{{d_{n - 1} }}{{d_n }}\phi _{n
- 1} (x) + \frac{{\lambda _{2n} }}{{2n}}(\beta _n  - x)\phi _n (x) = 0
\end{eqnarray*}

\noindent iii)  Raising operator
\begin{eqnarray*}
\fl ({\rm ND}3)\quad L^ +  (x,n) \equiv  \left[ {\frac{{\lambda _n }}{n}\frac{{\tau _n
(x)}}{{\tau '_n }} - \sigma (x)} \right]\phi _n (x) + \sqrt {\left( {\sigma (x - 1) + \tau
(x - 1)} \right)\sigma (x)} \phi _n (x - 1)  \\
\lo = \frac{{\lambda _{2n} }}{{2n}}\alpha _n
\frac{{d_{n + 1} }}{{d_n }}\phi _{n + 1} (x)
\end{eqnarray*}

\noindent iv)  Lowering operator
\begin{eqnarray*}
 \fl ({\rm ND}4)\quad L^ -  (x,n)\equiv \left[ { - \frac{{\lambda _n }}{n}\frac{{\tau _n
(x)}}{{\tau '_n }} + \lambda _n  + \frac{{\lambda _{2n} }}{{2n}}(x - \beta _n ) - \sigma
(x) - \tau (x)} \right]\phi _n (x)   \\ 
  + \sqrt {\left( {\sigma (x) + \tau (x)} \right)\sigma (x + 1)} \phi _n (x + 1) =
\frac{{\lambda _{2n} }}{{2n}}\gamma _n \frac{{d_{n - 1} }}{{d_n }}\phi _{n - 1} (x) 
\end{eqnarray*}

From the last two expressions we get all the solution of the difference equation
(ND1). Putting $n = 0$ in (ND4) we obtain $\phi _0 (x)$ and inserting this value en (DC3)
we obtain, by iteration, all the normalized function $\phi _n (x)$.

The explicit calculations for all the orthonormal functions of hypergeometric type are
given in table II.


As in section 3, we make two observations. Firstly, the raising and lowering operators
(ND3) and (ND4) are mutually adjoint in the case of Krauvchuk, Meixner and Charlier
functions. For the Hahn and Chebyshev functions we have to divide both by ${{\lambda
_{2n} } \mathord{\left/  {\vphantom {{\lambda _{2n} } {2n}}} \right. \kern-\nulldelimiterspace}
{2n}}$, therefore, they become mutually adjoint, namely,

\[
 \fl \sum\limits_{x = a}^{b - 1} {\phi _{n + 1} (x)\left[ {\frac{{2n}}{{\lambda _{2n} }}L^
+  (x,n)\phi _n (x)} \right]}   = \sum\limits_{x = a}^{b - 1} {\left[ {\frac{{2n +
2}}{{\lambda _{2n + 2} }}L^ -  (x,n + 1)\phi _{n + 1} (x)} \right]} \phi _n (x) = \alpha
_n \frac{{d_{n + 1} }}{{d_n }} 
\]

Secondly, the operator corresponding to the eigenvaue $\lambda _n $ in (ND1) is self adjoint. In
order to prove it, it is enough to show
\begin{eqnarray*}
\fl \sum\limits_{x = a}^{b - 1} {\phi _l (x)} \left\{ {\sqrt {\left( {\sigma (x) + \tau
(x)}\right) \sigma (x + 1)} \phi _n (x + 1)}\right. + \left. {\sqrt {\left( {\sigma (x - 1) + \tau (x
- 1)}\right)\sigma (x) } \phi _n (x - 1)} \right\} \\[-.5em]
\lo = \sum\limits_{x = a}^{b - 1} {\phi _n (x)} \left\{ {\sqrt {\left( {\sigma (x - 1) +
\tau (x - 1)}\right)\sigma (x) } \phi _l (x - 1) }\right. \\
+ \left. \sqrt {\left( {\sigma (x) + \tau (x)}\right)\sigma (x + 1) } \phi _l (x + 1)\right\}
\end{eqnarray*}

From the orthogonality conditions $\sigma (a) = \sigma (b) = 0$, we can write
\begin{eqnarray*}
\fl  \sum\limits_{x = a }^{b - 1}{{\phi _n (x)} \sqrt {\left( {\sigma (x - 1) + \tau (x-
1)}\right)\sigma (x) } \phi _l (x -1) } \\[-.5em] 
\lo =\sum\limits_{x' = a - 1}^{b - 2} {\phi _n (x' + 1)} \sqrt {\left( {\sigma (x') + \tau
(x')} \right) \sigma (x' + 1) } \phi _l (x') \\
\lo =  \sum\limits_{x = a}^{b - 1} {\phi _n (x +
1)} \sqrt {\left( {\sigma (x) + \tau (x)} \right) \sigma (x + 1) } \phi _l (x) 
\end{eqnarray*}

Similarly
\begin{eqnarray*}
\fl  \sum\limits_{x = a}^{b - 1} {{\phi _n (x)} \sqrt {\left( {\sigma (x) + \tau (x) } \right)
\sigma (x + 1)} \phi _l (x + 1)} \\[-.5em]
\lo = \sum\limits_{x = a}^{b - 1} {\phi _n (x - 1)} \sqrt
{\left( {\sigma (x - 1) + \tau (x - 1)} \right) \sigma (x) } \phi _l (x) 
\end{eqnarray*}

\section{Factorization for differential equation of hypergeometric type}

The raising and lowering operators of section 2 and 4 will help us to factorize the
second order differential equation of hypergeometric type into the product of two first
order operators in agreement with the general method of Infeld and Hull [1]. 

From (NC1) we define the operator 
\[H(s,n) \equiv \sigma (s)\frac{{d^2 }}{{ds^2 }} + \sigma
'(s)\frac{d}{{ds}} -
\frac{1}{4}\frac{{\left( {\sigma (s) - \sigma '(s)} \right)^2 }}{{\sigma (s)}} -
\frac{1}{2}(\tau ' - \sigma '') + \lambda _n \]
that satisfies $H(s,n)\psi _n (s) = 0$

We write the raising and lowering operators, (NC3) and (NC4) respectively, in the
following way 
\begin{eqnarray*}
 L^ +  (s,n) \equiv f(s,n) - \sigma (s)\frac{d}{{ds}}\\
L^ -  (s,n) \equiv g(s,n) + \sigma (s)\frac{d}{{ds}}
\end{eqnarray*}
where
\begin{eqnarray*}
f(s,n) = \frac{{\lambda _n }}{n}\frac{{\tau _n (s)}}{{\tau '_n }} + \frac{1}{2}\left(
{\tau (s) - \sigma '(s)} \right)\\
g(s,n) =  - \frac{{\lambda _n }}{n}\frac{{\tau _n (s)}}{{\tau '_n }} + \frac{{\lambda
_{2n} }}{{2n}}\left( {s- \beta _n } \right) - \frac{1}{2}\left( {\tau (s) - \sigma '(s)}
\right)
\end{eqnarray*}
satisfying
\[f(s,n - 1) = g(s,n)\quad {\rm or}\quad f(s,n) = g(s,n + 1)\]
which can be proved by Taylor expansion.

Now we calculate
\begin{eqnarray*}
\fl  L^ -  (s,n + 1)L^ +  (s,n) =  g(s,n + 1)f(s,n) + \sigma (s)\left\{ {f(s,n) - g(s,n +
1)}\right\} \frac{d}{{ds}}  \\  
   + \sigma (s)\left\{ {f'(s,n) - \sigma '(s)\frac{d}{{ds}} - \sigma (s)\frac{{d^2
}}{{ds^2 }}} \right\} 
\end{eqnarray*}

The second term of the right side becomes zero. Substituting the values for
$f(s,n)$, $g(s,n)$ and $H(s,n)$ we get
\begin{eqnarray*}
\fl L^ -  (s,n + 1)L^ +  (s,n) = \left[ {\left( {\frac{{\lambda _n }}{n}} \right)^2 \left(
{\frac{{\tau _n (s)}}{{\tau '_n }}} \right)^2  + \frac{{\lambda _n }}{n}\frac{{\tau _n
(s)}}{{\tau '_n }}\left( {\tau (s) - \sigma '(s)} \right) + (n + 1)\frac{{\lambda _n
}}{n}\sigma (s)} \right] \\
- \sigma (s)H(s,n)
\end{eqnarray*}

It can be proved that the expresion in squared brackets is independent of $s$, say $\mu
(n)$.  Applying the last equality to the orthonormal functions $\psi _n (s)$ and taking
into account (NC3) and (NC4) we get
\[\mu (n) = \frac{{\lambda _{2n} }}{{2n}}\frac{{\lambda _{2n + 2} }}{{2n + 2}}\alpha _n
\gamma _{n + 1}  \]

With the same technique we calculate
\begin{eqnarray*}
\fl L^ +  (s,n - 1)L^ -  (s,n) =  f(s,n - 1)g(s,n) + \sigma (s)\left\{ {f(s,n - 1) -
g(s,n)} \right\}\frac{d}{{ds}} -  \\  
   - \sigma (s)\left\{ {g'(s,n) + \sigma '(s)\frac{d}{{ds}} + \sigma (s)\frac{{d^2
}}{{ds^2 }}} \right\} \\  
\end{eqnarray*}

From the properties between $f(s,n)$ and $g(s,n)$, the second term in the right side becomes zero.
Substituting the values of these functions and $H(s,n)$ we finally obtain
\begin{eqnarray*}
\fl L^ +  (s,n - 1)L^ -  (s,n) =  \left[ {\left( {\frac{{\lambda _{n - 1} }}{{n - 1}}}
\right)^2 \left( {\frac{{\tau _{n - 1} (s)}}{{\tau '_{n - 1} }}} \right)^2  +
\frac{{\lambda _{n - 1} }}{{n - 1}}\frac{{\tau _{n - 1} (s)}}{{\tau '_{n - 1} }}\left(
{\tau (s) - \sigma ' (s)} \right)} \right. \\
+  \left. n\frac{{\lambda _{n - 1} }}{{n - 1}}\sigma (s)
\right] -\sigma (s)H(s,n)
\end{eqnarray*}

It can be proved that the expresion in squared bracket is independent of $s$, say $\nu (n)$.
Applying the last equality to the orthonormal functions $\psi _n (s)$ and taking into account
(NC3) and (NC4) we get
\[\nu (n) = \frac{{\lambda _{2n - 2} }}{{2n - 2}}\frac{{\lambda _{2n} }}{{2n}}\alpha _{n
- 1} \gamma _n \]

Obviously, $\nu (n + 1) = \mu (n)$. These constants are given explicitely in Table I

Finally we have the desired relation equivalent to the Infeld-Hull-Miller factorization method:
\begin{eqnarray*}
\fl ({\rm NC}5)\qquad \quad  L^ -  (s,n + 1)L^ +  (s,n) = \mu (n) - \sigma (s)H(s,n) \\[.5em]
\fl ({\rm NC}6)\qquad \quad L^ +  (s,n)L^ -  (s,n + 1) = \mu (n) - \sigma (s)H(s,n + 1)
\end{eqnarray*}

If we want $L^ +  (s,n)$ and $L^ -  (s,n)$ mutually adjoint we have to divide both sides
of (NC5) and (NC6) by 
\[\frac{{\lambda _{2n + 2} }}{{2n + 2}}\frac{{\lambda _{2n} }}{{2n}}\]

\section{Factorization of difference equation of hypergeometric type}

For the case of orthonormal hypergeometric functions of discrete variable, we define 
from (ND1) the operator  
\begin{eqnarray*}
\fl H(x,n) \equiv \sqrt {\left( {\sigma (x) + \tau (x)} \right)\sigma (x + 1)} E^ +   +
\sqrt {\left( {\sigma (x - 1) + \tau (x - 1)} \right)\sigma (x)} E^- \\  - \left(
{2\sigma (x) + \tau (x)} \right) + \lambda _n 
\end{eqnarray*}
 where $E^ +  f(x) = f(x + 1),\quad E^ - f(x) = f(x - 1)$, and the orthonormal functions
satisfy 
\[H(x,n)\phi _n (x) = 0\]

As before we write the raising and lowering operators in the following way
\begin{eqnarray*}
 L^ +  (x,n)  = u(x,n) + \sqrt {\left( {\sigma (x - 1) + \tau (x - 1)} \right)\sigma
(x)} E^-  \\[.5em]
L^ -  (x,n)  = v(x,n) + \sqrt {\left( {\sigma (x) + \tau (x)} \right)\sigma (x + 1)} E^+ 
\end{eqnarray*}
where
\begin{eqnarray*}
 u(x,n)  = \frac{{\lambda _n }}{n}\frac{{\tau _n (x)}}{{\tau '_n }} - \sigma (x) \\[.5em]
v(x,n)  =  - \frac{{\lambda _n }}{n}\frac{{\tau _n (x)}}{{\tau '_n }} + \lambda _n  +
\frac{{\lambda _{2n} }}{{2n}}(x  - \beta _n ) - \sigma (x) - \tau (x) 
\end{eqnarray*}

Both expressions satisfy
\begin{eqnarray*}
u(x + 1,n) = v(x,n + 1)\quad {\rm or} \\[.5em]
u(x + 1,n - 1) = v(x,n)
\end{eqnarray*}
that can  be proved by Taylor expansion.

Now we calculate 
\begin{eqnarray*}
\fl  L^ -  (x,n + 1)  L^ +  (x,n) = v(x,n + 1)u(x,n) + \left( {\sigma (x) + \tau (x)}
\right)\sigma (x + 1)+u(x + 1,n)  \\[.2em]  
  \times \left\{ {\sqrt {\left( {\sigma (x) + \tau (x)} \right)\sigma (x + 1)} E^
+   + \sqrt {\left( {\sigma (x - 1) + \tau (x - 1)} \right)\sigma (x)} E^ -  } \right\}
\end{eqnarray*}

Substituting the values for $u(x,n), v(x,n)$ and $H(x,n)$ we get
\begin{eqnarray*}
\fl L^ -  (x,n + 1)  L^ +  (x,n) =  \left[ {\left( {\frac{{\lambda _n }}{n}\frac{{\tau _n
(x)}}{{\tau '_n }} - \lambda _n } \right)\left( {\frac{{\lambda _n }}{n}\frac{{\tau _n
(x + 1)}}{{\tau '_{n + 1} }} - \sigma (x + 1)} \right)}\right.  \\
 + \left.  {\frac{{\lambda _n}}{n}\frac{{\tau _n (x + 1)}}{{\tau '_n }}\left( {\sigma (x)
 +  \tau (x)} \right)} \right] + u(x + 1,n)H(x,n) 
\end{eqnarray*}

It can be proved that the expresion in squared bracket is independent of $x$, say $\mu (n)$.
Applying the last equality to the orthonormal function $\phi _n (x)$
 and taking into account (ND1), (ND3) and (ND4) we get
\[\mu (n) =   \frac{{\lambda _{2n} }}{{2n}}\frac{{\lambda _{2n + 2} }}{{2n + 2}}\alpha
_n \gamma _{n + 1} \]

With the same technique we calculate
\begin{eqnarray*}
\fl  L^ +  (x,n - 1)L^ -  (x,n) = \left[ \left( { - \frac{{\lambda _n
}}{n}\frac{{\tau _n (x - 1)}}{{\tau '_n }} +\frac{{\lambda _{2n} }}{{2n}}(x - 1 - \beta _n ) +
\lambda _n }
\right)
\right. \\
  \times  \left( { -\frac{{\lambda _n }}{n}\frac{{\tau _n (x)}}{{\tau '_n }} +
\frac{{\lambda _{2n} }}{{2n}}(x - \beta _n ) + \sigma (x)} \right) - \left( {\sigma (x - 1) +
\tau (x - 1)} \right) \\
\times \left. \Big( - \frac{\lambda _n}{n}\frac{\tau _n (x)}{\tau '_n} + \frac{\lambda_{2n}}
{2n}\left( {x - \beta _n }\Big) \right) \right] + u(x,n - 1)H(x,n) 
\end{eqnarray*}

As before the expression in squared brackets is independent of $x$, say $\nu (n)$. Applying both
sides of the last equality to the functions $\phi _n (x)$, and taking into account (ND1) (ND3)
and (ND4) we obtain
\[\nu (n) = \frac{{\lambda _{2n - 2} }}{{2n - 2}}\frac{{\lambda _{2n} }}{{2n}}\alpha _{n
- 1} \gamma _n \] 

Obviously $\nu (n + 1) = \mu (n)$

These constants are given explicitely in Table II.

Finally the desired relations corresponding to the Spiridonov-Vinet-Zhedanov factorization method are
\begin{eqnarray*}
\fl ({\rm ND}5)\qquad \quad  L^ -  (x,n + 1)L^ +  (x,n) = \mu (n) + u(x + 1,n)H(x,n) \\[.5em]
\fl ({\rm ND}6)\qquad \quad  L^ +  (x,n)L^ -  (x,n + 1) = \mu (n) + u(x,n - 1)H(x,n + 1)
\end{eqnarray*}

Again, if we want $L^ +  (x,n)$ and $L^ -  (x,n)$ to be mutually adjoint, we have to divide both
expressions (ND5) and (ND6)by
\[\frac{{\lambda _{2n} }}{{2n}}\frac{{\lambda _{2n + 2} }}{{2n + 2}}\]
only in the case of Hahn and Chebyshev functions.

\section{Some comments}

The classical orthogonal polynomials we have presented in the preceeding sections are solutions of the second order
differential equation  

\[\sigma (s)y''_n(s)+\tau (s)y'_n(s)+\lambda _ny_n(s)=0\]

in the continuous case, or second order difference equation

\[\sigma (x)\Delta \nabla y_n(x)+\tau (x)\Delta y_n(x)+\lambda _ny_n(x)=0\]
in the discrete case for uniform lattices where $\sigma (x)$ and $\tau (x)$ are polynomials of at most the second and
first degree respectively .

Atakishiev and collaborators have generalized the classical orthogonal polynomials using a characterization based on the
difference equation of hypergeometric type that covers all the cases defined by Andrews and Askey [16]. This
characterization covers the q-analogue of classical orthogonal polynomials on non-uniform lattices. 

Our paper should be implemented with the construction of raising and lowering operators for the orthogonal polynomials on
non-uniform lattices in particular the q-analogue of the classical orthogonal polynomials. For this purpose we have at our
disposal the analogue of difference equations, Rodrigues formula, recurrence relations for the orthogonal polynomials on
non-uniform lattice, given explicetely by Nikiforov, Suslov and Uvarov [4].

An other approach for the same problem is given by Smirnov, via the factorization method suggested by Schr\"odinger for the
solution of second order differential equation of hypergeometric type. Smirnov has applied this method to the finite
difference equation on uniform lattices [9] and on non-uniform lattices [17], [18]. In his approach the raising and
lowering operators are defined with respect to two indices: the first one, the degree of the orthogonal polynomials, the
second one the order of the finite derivative with respect to the discrete variable. For this reason his raising and
lowering operators are not equal to ours.

A final comment to Tables I and II. In an unpublished Report of R. Koekoek and R.F. Swarttouw [20] tables are presented
for orthogonal polynomials of the Askey-scheme and its q-analogue; among them one finds the raising and lowering operators
of classical orthogonal polynomials of hypergeometric type. There are two points by which our tables are different from
theirs. First, we have calculated the raising and lowering operators from Rodrigues formula (see (1, 2, 13) and (2, 2, 10)
of Ref. [4]), but their raising and lowering operators are connected with some recurrence relations (see (1, 4, 5) and (2,
4, 13-17) of Ref. [4]) which are defined with respect to two indices. Besides that their tables do not cover the
differential/difference equations, recurrence relations and raising/lowering operators with respect to the orthonormal
functions of hyper geometric type as given in our tables.

\ack{
The author wants to express his gratitude to Prof. Smirnov for valuable conversations and
to Prof. A Ronveaux for his advise and encouragement and for bringing to his attention the papers of
A.P. Magnus and G. Bangerezako. He is also very thankful to the Referees for their suggestions and new references. This
work has been partially supported by D.G.I.C.Y.T. under contract \#Pb96-0538 (Spain).}

\newpage

\begin{center}
{\bf TABLE I}

{\bf Orthogonal polynomials of continuous variable}
\end{center}

\bigskip
\noindent {\sl Hermite polynomials}
\begin{eqnarray*}
\fl {\rm He 1)} \quad  H''_n (s) - 2sH'_n (s) + 2nH_n (s) = 0 \\
\fl {\rm He 2)} \quad  sH_n (s) = \frac{1}{2}H_{n + 1} (s) + nH_{n - 1} (s) \\
\fl {\rm He 3)} \quad  H_{n + 1} (s) = 2sH_n (s) - H'_n (s) \\
\fl {\rm He 4)} \quad  H_{n - 1} (s) = \frac{1}{{2n}}H'_n (s)
\end{eqnarray*}

\medskip
\noindent {\sl Laguerre polynomials}
\begin{eqnarray*}
\fl {\rm La 1)}\quad sL_n^{\alpha ''} (s) + (1 + \alpha  - s)L_n^{\alpha '} (s) +
nL_n^\alpha  (s) = 0 \\
\fl {\rm La 2)}\quad (n + 1)L_{n + 1}^\alpha  (s) + (n + \alpha )L_{n - 1}^\alpha  (s) + (s
- 2n - \alpha  - 1)L_n^\alpha  (s) = 0 \\
\fl {\rm La 3)}\quad (n + 1)L_{n + 1}^\alpha  (s) = (s - n - \alpha  - 1)L_n^\alpha  (s) +
sL_n^{\alpha '} (s) \\
\fl {\rm La 4)}\quad (n + \alpha )L_{n - 1}^\alpha  (s) = nL_n^\alpha  (s) - sL_n^{\alpha '}
(s)
\end{eqnarray*}

\medskip
\noindent {\sl Legendre polynomials}
\begin{eqnarray*}
\fl {\rm Le 1)}\quad  (1 - s^2 )P''_n (s) - 2sP'_n (s) + n(n + 1)P_n (s) = 0 \\
\fl {\rm Le 2)}\quad  \frac{{n + 1}}{{2n + 1}}P_{n + 1} (s) + \frac{n}{{2n + 1}}P_{n - 1} (s)
- sP_n (s) = 0 \\
\fl {\rm Le 3)}\quad  (n + 1)P_{n + 1} (s) = (n + 1)sP_n (s) - (1 - s^2 )P'_n (s)\\
\fl {\rm Le 4)}\quad  nP_{n - 1} (s) = nsP_n (s) + (1 - s^2 )P'_n (s)
\end{eqnarray*}

\medskip
\noindent {\sl Jacobi polynomials}
\begin{eqnarray*}
\fl {\rm J 1)}\quad  (1 - s^2 )P_n^{(\alpha ,\beta )^{\prime \prime}}
(s) + \left[ {\beta  - \alpha  - (\alpha  + \beta  + 2)s} \right]P_n^{(\alpha ,\beta ) ^\prime}(s)\\[0.5em]
 \lo + n(n +\alpha  + \beta  + 1)P_n^{(\alpha ,\beta )} (s) = 0  \\[1em] 
\fl {\rm J 2)}\quad  \frac{{2(n + 1)(n + \alpha  + \beta  + 1)}}{{(2n + \alpha  + \beta  + 1)(2n +
\alpha  + \beta  + 2)}}P_{n + 1}^{(\alpha ,\beta )} (s) \\[0.5em]
\lo + \frac{{2(n + \alpha )(n +
\beta )}}{{(2n + \alpha  + \beta )(2n + \alpha  + \beta  + 1)}}P_{n - 1}^{(\alpha ,\beta )} (s) 
\\[0.5em]
\lo + \left[ {\frac{{\beta ^2  - \alpha ^2 }}{{(2n + \alpha  + \beta )(2n + \alpha  +
\beta  + 2)}} - s} \right]P_n^{(\alpha ,\beta )} (s) = 0 \\[1em]
\fl {\rm J 3)}\quad \frac{{2(n + 1)(n + \alpha  + \beta  + 1)}}{{(2n + \alpha  + \beta  +
2)}}P_{n + 1}^{(\alpha ,\beta )} (s) \\[0.5em]
\lo =\left[ {\frac{{(n + \alpha  + \beta  + 1)}}{{(2n +
\alpha  + \beta  + 2)}}(\alpha  - \beta ) + (n + \alpha  + \beta  + 1)s}
\right] P_n^{(\alpha ,\beta )} (s) \\[0.5em]
\lo  - (1 - s^2 )P_n^{(\alpha ,\beta ) ^\prime}  (s) \\[1em]
\fl {\rm J 4)}\quad  \frac{{2(n + \alpha )(n + \beta )}}{{(2n + \alpha  + \beta )}}P_{n -
1}^{(\alpha ,\beta )} (s) = \Bigg[ \frac{{(n + \alpha  + \beta  + 1)}}{{(2n + \alpha  + \beta 
+ 2)}}(\beta  - \alpha ) - {(n + \alpha  + \beta  + 1)s }\\[0.5em]
\lo +   (2n + \alpha  + \beta  + 1)\left( {s -\frac{{\beta ^2  - \alpha ^2 }}{{(2n + \alpha  + \beta )(2n + \alpha  + \beta  + 2)}}}
\right) \Bigg]P_n^{(\alpha ,\beta ) ^\prime}  (s)  \\[0.5em]
\lo + (1 - s^2 )P_n^{(\alpha ,\beta )} (s)
\end{eqnarray*}

\medskip
\noindent {\sl Normalized Hermite functions}
$$\psi _n (s) = \left( {2^n n!\sqrt \pi  } \right)^{ - {1 \mathord{\left/
 {\vphantom {1 2}} \right.\kern-\nulldelimiterspace} 2}} e^{ - {{s^2 } \mathord{\left/
 {\vphantom {{s^2 } 2}} \right.
 \kern-\nulldelimiterspace} 2}} H_n (s)$$
\begin{eqnarray*}
\fl {\rm N He 1)}\quad  \psi ''_n (s) + (1 - s^2 )\psi _n (s) + 2n\psi _n (s) = 0 \\
\fl {\rm N He 2)}\quad  \sqrt {2(n + 1)} \psi _{n + 1} (s) + \sqrt {2n} \psi _{n - 1} (s) -
2s\psi _n (s) = 0 \\
\fl {\rm N He 3)}\quad  L^ +  (s,n)\psi _n (s) = s\psi _n (s) - \psi '_n (s) = \sqrt {2(n + 1)}
\psi _{n + 1} (s) \\ 
\fl  {\rm N He 4)}\quad  L^ -  (s,n)\psi _n (s) = s\psi _n (s) + \psi '_n (s) = \sqrt {2n}
\psi _{n - 1} (s) \\ 
\lo\psi _0 (s) = \pi ^{ - {1 \mathord{\left/
 {\vphantom {1 4}} \right.
 \kern-\nulldelimiterspace} 4}} e^{{{ - s^2 } \mathord{\left/
 {\vphantom {{ - s^2 } 2}} \right.
 \kern-\nulldelimiterspace} 2}}  \\ 
\lo \psi _n (s) = \frac{1}{{\sqrt {2^n n!} }}\left( {s - \frac{d}{{ds}}} \right)^n \psi
_0 (s) \\ 
\lo L^ +  (s,n)L^ -  (s,n) = 2n\psi _n (s) \\ 
\lo L^ -  (s,n)L^ +  (s,n)\psi _n (s) = 2(n + 1)\psi _n (s) 
\end{eqnarray*}
 
\medskip
\noindent {\sl Normalized Laguerre functions}
$$\psi _n (s) = \sqrt {\frac{{n!}}{{\Gamma (n + \alpha  + 1)}}} e^{{{ - s} \mathord{\left/
 {\vphantom {{ - s} 2}} \right.
 \kern-\nulldelimiterspace} 2}} s^{{\alpha  \mathord{\left/
 {\vphantom {\alpha  2}} \right.
 \kern-\nulldelimiterspace} 2}} L_n^\alpha  (s)$$
\begin{eqnarray*}
\fl {\rm N La 1)}\quad  s\psi ''_n (s) + \psi '_n (s) - \frac{1}{4}\left( {s + \frac{{\alpha
^2 }}{s} - 2\alpha  - 2} \right)\psi _n (s) + n\psi _n (s) = 0 \\
\fl {\rm N La 2)}\quad \sqrt {(n + 1)(n + \alpha  + 1)} \psi _{n + 1} (s) + \sqrt {n(n
+ \alpha )} \psi _{n - 1} (s) - (2n + \alpha  + 1 - s)\psi _n (s) = 0 \\
\fl {\rm N La 3)}\quad  L^ +  (s,n)\psi _n (s) =  - \frac{1}{2}(2n + \alpha  + 2 -
s)\psi _n (s) - s\psi '_n (s) \\[0.3em]
\lo = - \sqrt {(n + 1)(n + \alpha  + 1)} \psi _{n + 1} (s) \\
\fl {\rm N La 4)}\quad  L^ -  (s,n)\psi _n (s) =  - \frac{1}{2}(2n + \alpha  - s)\psi _n
(s) + s\psi '_n (s) = - \sqrt {n(n + \alpha )} \psi _{n - 1} (s) \\ 
  \lo \psi _0 (s) = \sqrt {\frac{1}{{\Gamma (\alpha  + 1)}}} e^{ - {s \mathord{\left/
 {\vphantom {s 2}} \right.
 \kern-\nulldelimiterspace} 2}} s^{{\alpha  \mathord{\left/
 {\vphantom {\alpha  2}} \right.
 \kern-\nulldelimiterspace} 2}}  \\ 
  \lo \psi _n (s) =  {\frac{1}{{\sqrt {n!(\alpha  + 1)_n } }}\prod\limits_{k = 0}^{n
- 1} {L^ +  (s,n - 1 - k)} } \psi _0 (s) \\ 
  \lo L^ +  (s,n - 1)L^ -  (s,n)\psi _n (s) = n(n + \alpha )\psi _n (s) \\ 
  \lo L^ -  (s,n + 1)L^ +  (s,n)\psi _n (s) = (n + 1)(n + \alpha  + 1)\psi _n (s)
\end{eqnarray*}

\medskip
\noindent {\sl Normalized Legendre functions}
$$\psi _n (s) = \sqrt {\frac{{2n + 1}}{2}} P_n (s)$$
\begin{eqnarray*}
\fl {\rm N Le 1)}\quad  (1 - s^2 )\psi ''_n (s) - 2s\psi '_n (s) + n(n + 1)\psi _n (s) = 0 \\
\fl {\rm N Le 2)}\quad  (n + 1)\sqrt {\frac{{2n + 1}}{{2n + 3}}} \psi _{n + 1} (s) + n\sqrt
{\frac{{2n + 1}}{{2n - 1}}} \psi _{n - 1} (s) - (2n + 1)s\psi _n (s) = 0 \\
\fl {\rm N Le 3)}\quad  L^ +  (s,n)\psi _n (s) = (n + 1)s\psi _n (s) - (1 - s^2 )\psi '_n (s)
= (n + 1)\sqrt {\frac{{2n + 1}}{{2n + 3}}} \psi _{n + 1} (s) \\
\fl {\rm N Le 4)}\quad  L^ -  (s,n)\psi _n (s) = ns\psi _n (s) + (1 - s^2 )\psi '_n (s) =
n\sqrt {\frac{{2n + 1}}{{2n - 1}}} \psi _{n - 1} (s) \\ 
 \lo\psi _0 (s) = \frac{1}{{\sqrt 2 }} \\ 
 \lo\psi _n (s) = \frac{1}{{n!}}\sqrt {2n + 1} \prod\limits_{k = 0}^{n - 1} {L^ +  (s,n
- 1 - k)\psi _0 (s)}  \\ 
 \lo L^ +  (s,n - 1)L^ -  (s,n) = n^2  \\ 
 \lo L^ -  (s,n + 1)L^ +  (s,n) = (n + 1)^2 
\end{eqnarray*}

\medskip
\noindent {\sl Normalized Jacobi functions}
$$\psi _n (s) = \sqrt {\frac{{n!(2n + \alpha  + \beta  + 1)(n + \alpha  + \beta  +
1)}}{{2^{\alpha  + \beta  + 1} \Gamma (n + \alpha  + 1)\Gamma (n + \beta  + 1)}}} (1 -
s)^{\frac{\alpha }{2}} (1 + s)^{\frac{\beta }{2}} P_n^{(\alpha ,\beta )} (s)$$
\begin{eqnarray*}
\fl {\rm NJ 1)}\quad  (1 - s^2 )\psi ''_n (s) - 2s\psi '_n (s) - \frac{1}{4}\left\{
{\frac{{\left( {\beta  - \alpha  - (\alpha  + \beta )s} \right)^2 }}{{1 - s^2 }} -
2(\alpha  + \beta )s} \right\}\psi _n (s) +\\
\lo + n(n + \alpha  + \beta  + 1)\psi _n (s) = 0 \\[1em]
\fl {\rm NJ 2)}\quad \frac{{2\sqrt {(n + 1)(n + \alpha  + 1)(n + \beta  + 1)(n + \alpha  +
\beta  + 1)(2n + \alpha  + \beta  + 1)} }}{{(2n + \alpha  + \beta  + 2)\sqrt {2n + \alpha 
+ \beta  + 3} }}\psi _{n + 1} (s) + \\[0.5em]
\lo  + \frac{{2\sqrt {n(n + \alpha )(n + \beta )(n + \alpha  + \beta )(2n + \alpha  + \beta 
+ 1)} }}{{(2n + \alpha  + \beta  + 2)\sqrt {2n + \alpha  + \beta  - 1} }}\psi _{n - 1} (s)
 + \\[0.2em]
\lo +\; (2n + \alpha  + \beta  + 1)\left\{ {\frac{{\beta ^2  - \alpha ^2 }}{{(2n +
\alpha  +
\beta )(2n + \alpha  + \beta  + 2)}} - s} \right\}\psi _n (s) = 0 \\[1em] 
\fl {\rm NJ 3)}\quad  L^ +  (s,n) \psi _n (s)  =\Bigg\{ (n + \alpha  + \beta  +
1)s  - \frac{{n + \alpha  + \beta  + 1}}{{2n + \alpha  + \beta  + 2}}(\beta  - \alpha 
- n^2 ) \\[0.2em]
\lo + \frac{1}{2}(\beta  - \alpha  - (\alpha  + \beta )s) \Bigg\}  \psi _n (s) - (1
- s^2 )\psi '_n (s) = \\[.5em]
\lo = \frac{{2\sqrt {(n + 1)(n + \alpha  + 1)(n + \beta  + 1)(n +
\alpha  +
\beta  + 1)(2n + \alpha  + \beta  + 1)} }}{{(2n + \alpha  + \beta  + 2)\sqrt {2n + \alpha 
+ \beta  + 3} }}\psi _{n + 1} (s)    \\[1em] 
\fl {\rm NJ 4)}\quad   L^ -  (s,n)\psi _n (s) =  \Bigg\{  - (n + \alpha  + \beta  + 1)s + \frac{{n +
\alpha  + \beta  + 1}}{{2n + \alpha  + \beta  + 2}}(\beta  - \alpha  - n^2 ) 
\\[0.2em]
    \lo +\;(2n +\alpha  + \beta  + 1)\left( {s - \frac{{\beta ^2  - \alpha ^2
}}{{(2n +\alpha  + \beta )(2n + \alpha  +
\beta  + 2)}}} \right) \\[0.2em]
\lo - \frac{1}{2}\left( {\beta  - \alpha  - (\alpha  + \beta )s}
\right) \Bigg\} \psi _n (s) + (1 - s^2 ) \psi '_n (s) \\[0.2em]
\lo = \frac{{2\sqrt {n(n + \alpha )(n + \beta )(n
+ \alpha  + \beta )(2n +
\alpha  + \beta  + 1)} }}{{(2n + \alpha  + \beta )\sqrt {2n + \alpha  + \beta  - 1} }}\psi
_{n - 1} (s) \\[1em] 
\lo \psi _0\left( s \right)={{\alpha +\beta +1} \over {\sqrt {2^{\alpha +\beta +1}\Gamma \left(
{\alpha +1} \right)\Gamma \left( {\beta +1} \right)}}}\left( {1-s} \right)^{{\alpha  \over
2}}\left( {1+s} \right)^{{\beta  \over 2}}\\[1em]
\lo \psi _n (s) = \prod\limits_{k = 0}^{n - 1} \Bigg\{ \frac{(2k + \alpha  + \beta  + 2)}
{{2\sqrt {(k + 1)(k + \alpha + 1)(k + \beta + 1)}}} \\[0.2em]
\qquad \qquad \times \frac{\sqrt {2k +
\alpha  + \beta  + 3}}{\sqrt {(k + \alpha  +\beta  + 1)(2k + \alpha + \beta + 1)}}\; L^ +  (s,n-1-k)\Bigg\}\psi _0 (s) \\[1em]
\lo L^ +  (s,n - 1)L^ -  (s,n)\psi _n (s) = \frac{{4n(n + \alpha )(n + \beta )(n + \alpha  +
\beta )}}{{(2n + \alpha  + \beta )^2 }}\psi _n (s) \\[1em]
\lo L^ -  (s,n + 1)L^ +  (s,n)\psi _n (s) \\[0.2em]
\qquad \qquad = \frac{{4(n + 1)(n + \alpha  + 1)(n + \beta  +
1)(n + \alpha  + \beta  + 1)}}{{(2n + \alpha  + \beta  + 2)^2 }}\psi _n (s)
\end{eqnarray*}


\ \newline

\begin{center}
{\bf TABLE II}

\medskip
\noindent {\bf  Orthogonal Polynomials of Discrete Variable}
\end{center}

\medskip
\noindent {\sl Kravchuk polynomials}
\begin{eqnarray*}
\fl {\rm K1)}\quad \frac{{p(N - x)}}{q}k_n (x + 1) + xk_n (x - 1) + \frac{{x(p - q) -
Np}}{q}k_n (x) + \frac{n}{q}k_n (x) = 0 \\
\fl {\rm K2)}\quad \frac{{n + 1}}{q}k_{n + 1} (x) + p(N - n + 1)k_{n - 1} (x) + \left[ {n +
p(N - 2n) - x} \right]k_n (x) = 0 \\
\fl {\rm K3)}\quad \frac{{n + 1}}{q}k_{n + 1} (x) = \frac{p}{q}(x + n - N)k_n (x) + xk_n (x -
1) \\
\fl {\rm K4)}\quad  p(N - n + 1) = \frac{p}{q}(x + n - N)k_n (x) + \frac{p}{q}(N - x)k_n (x +
1)
\end{eqnarray*}

\medskip
\noindent {\sl Meixner polynomyals}
\begin{eqnarray*}
\fl {\rm M1)}\quad  \mu (x + \gamma )m_n (x + 1) + xm_n (x - 1) - \left[ {\mu (x + \gamma ) +
x} \right]m_n (x) + n(1 - \mu )m_n (x) = 0 \\[0.5em]
\fl {\rm M2)}\quad  \mu m_{n + 1} (x) - n(n + \gamma  - 1)m_{n + 1} (x) + \left[ {\mu (x + n +
\gamma ) + n - x} \right]m_n (x) = 0 \\[0.5em]
\fl {\rm M3)}\quad   - \mu m_{n + 1} (x) =  - \mu (x + n + \gamma )m_n (x) + xm_n (x - 1)
\\[0.5em]
\fl {\rm M4)}\quad   - n(n + \gamma  - 1)m_{n - 1} (x) =  - \mu (x + n + \gamma
)m_n (x) + \mu (x + \gamma )m_n (x + 1)
\end{eqnarray*}

\bigskip
\noindent {\sl Charlier polynomials}
\begin{eqnarray*}
\fl {\rm C1)}\quad  \mu c_n (x + 1) + xc_n (x - 1) - (x + \mu )c_n (x) + nc_n (x) = 0
\\[0.5em] 
\fl {\rm C2)}\quad  - \mu c_{n + 1} (x) - nc_{n - 1} (x) + (n + \mu  - x)c_n (x) = 0
\\[0.5em] 
\fl {\rm C3)}\quad  - \mu c_{n + 1} (x) =  - \mu c_n (x) + xc_n (x - 1) \\[0.5em]
\fl {\rm C4)}\quad  - nc_{n - 1} (x) =  - \mu c_n (x) + \mu c_n (x + 1)
\end{eqnarray*}

\bigskip
\noindent {\sl Chebyshev polynomials}
\begin{eqnarray*}
\fl {\rm T1)}\quad (x + 1)(N - x - 1)t_n (x + 1) + x(N - x)t_n (x - 1) - \\
\lo -\ \left[ {(N - x -
1)(x + 1) + x(N - x)} \right]t_n (x) + n(n + 1)t_n (x) = 0\\[0.5em]
\fl {\rm T2)}\quad \frac{1}{2}(n + 1)t_{n + 1} (x) + \frac{1}{2}n\left( {N^2  - n^2 }
\right)t_{n - 1} (x) + \frac{1}{2}(2n + 1)\left( {N - 1 - 2x} \right)t_n (x) = 0\\[0.5em]
\fl {\rm T3)}\quad \frac{1}{2}(n + 1)t_{n + 1} (x) \\[0.2em]
\lo =  - \left[ {\frac{1}{2}(n + 1)(N - 2x - n
- 1) + x(N - x)} \right]t_n (x) + x(N - x)t_n (x - 1)\\[0.5em]
\fl {\rm T4)}\quad \frac{1}{2}n\left( {N^2  - n^2 } \right)t_{n - 1} (x) \\[0.2em]
\lo = \Bigg[
{\frac{1}{2}(n + 1)(N - 2x - n - 1) + n(n + 1) + (2n + 1)\left( {x - \frac{{N - 1}}{2}}
\right)} \\
{\lo - \ (x + 1)(N - x - 1)} \Bigg]t_n (x) + (x + 1)(N - x - 1)t_n (x + 1)
\end{eqnarray*}

\bigskip
\noindent {\sl Hahn polynomials}
\begin{eqnarray*}
\fl {\rm Ha1)}\quad \left[ {x(N - x - \beta  - 2) + (\beta  + 1)(N - 1)} \right]h_n^{\alpha
,\beta } (x + 1) + x(N + \alpha  - x)h_n^{\alpha ,\beta } (x - 1) \\[0.2em]
\lo  - \left[ {x(2N - 2x + \alpha  - \beta  - 2) + (\beta  + 1)(N - 1)} \right]h_n^{\alpha
,\beta } (x) \\[0.2em]
\lo  + n(n + \alpha  + \beta  + 1)h_n^{\alpha ,\beta } (x) = 0 \\[1em]
\fl {\rm Ha2)}\quad \frac{{(n + 1)(n + \alpha  + \beta  + 1)}}{{2n + \alpha  + \beta  +
2}}h_{n + 1}^{\alpha ,\beta } (x) \\[0.2em]
\lo + \frac{{(n + \alpha )(n + \beta )(N + n + \alpha  +
\beta )(N - n)}}{{2n + \alpha  + \beta }}h_{n - 1}^{\alpha ,\beta }  +  (2n + \alpha  + \beta  + 1) \\[0.2em]
\times \Big[{\frac{{\alpha  - \beta  + 2N - 2}}{4} +
\frac{{\left( {\beta ^2  - \alpha ^2 } \right)(2N + \alpha  + \beta )}}{{4(2n + \alpha  +
\beta )(2n + \alpha  + \beta  + 2)}} - x} \Big]h_n^{\alpha ,\beta } (x) = 0 \\[1em]
\fl  {\rm Ha3)}\quad \frac{{(n + 1)(n + \alpha  + \beta  + 1)}}{{2n + \alpha  + \beta  +
2}}h_{n + 1}^{\alpha ,\beta } (x) = x(N + \alpha  - x)h_n^{\alpha ,\beta } (x - 1) \\
\lo -\Big\{   \frac {{(n + \alpha  + \beta  + 1)}}{{2n + \alpha  + 
\beta  + 2}}\Big[ (\beta  + 1)(N - 1) - (\alpha  + \beta  + 2 + 2n)x  \\
  \lo  + (N - n - \beta  - 2)n \Big] + x(N + \alpha  - x) \Big\}h_n^{\alpha ,\beta } (x) \\[1em]
\fl {\rm Ha4)}\quad \frac{{(n + \alpha )(n + \beta )(N + n + \alpha  + \beta )(N - n)}}{{2n +
\alpha  + \beta}}h_{n - 1}^{\alpha ,\beta } (x)  = \\[.5em]
   \lo = \Big[ x(N - x - \beta  - 2) + (\beta  + 1)(N - 1) \Big] h_n^{\alpha ,\beta } (x + 1) \\[0.2em]
+ \Bigg[ \frac{{n +
\alpha  + \beta  + 1}}{{2n + \alpha  + \beta  + 2}} \Big( (\beta  + 1)(N - 1) - (\alpha  +
\beta  + 2 + 2n)x \\[0.2em]
+ (N - n - \beta  - 2)n \Big) + n(n + \alpha  + \beta  + 1) + (2n + \alpha  + \beta  + 1) \\[0.2em]
\times \Big( x -\frac{{\alpha  - \beta  + 2N - 2}}{4}  - \frac{{\left( {\beta ^2  - \alpha ^2 } \right)(2N +
\alpha  + \beta )}}{{4(2n + \alpha  + \beta )(2n + \alpha  + \beta  + 2)}} \Big) \\[0.2em]
-  \ x(N - x - \beta  - 2) + (\beta  + 1)(N - 1) \Bigg] h_n^{\alpha ,\beta } (x)
\end{eqnarray*}

\medskip
\noindent {\sl Normalized Kravchuk functions}
$$\quad \psi _n (x) = \sqrt {\frac{{n!(N - n)!}}{{(pq)^n }}} \sqrt {\frac{{p^x q^{N - x}
}}{{x!(N - x)!}}} k_n (x)$$
\begin{eqnarray*}
\fl {\rm NK1)}\quad \sqrt {\frac{p}{q}(N - x)(x + 1)} \psi _n (x + 1) + \sqrt {\frac{p}{q}(N
- x + 1)x} \psi _n (x - 1)  
  + \\
\lo \qquad +\frac{{x(p - q) - Np}}{q}\psi _n (x) + \frac{n}{q}\psi _n (x) = 0 \\[1em] 
\fl {\rm NK2)}\quad \sqrt {\frac{p}{q}(N - n)(n + 1)} \psi _{n + 1} (x) + \sqrt {\frac{p}{q}(N
- n + 1)n} \psi _{n - 1} (x) \\[0.2em]
  + \frac{1}{q}\left[ {n + p(N - 2n) - x} \right]\psi _n (x)=0 \\[1em]
\fl {\rm NK3)}\quad L^ +  (x,n)\psi _n (x) = \frac{p}{q}(x + n - N)\psi _n (x) + \sqrt {\frac{p}{q}(N -
x + 1)x} \psi _n (x - 1) \\[0.2em]
= \sqrt {\frac{p}{q}(N - n)(n + 1)} \psi _{n + 1} (x) \\[1em]
\fl {\rm NK4)}\quad L^ -  (x,n)\psi _n (x) = \frac{p}{q}(x + n - N)\psi _n (x) + \sqrt {\frac{p}{q}(N -
x)(x + 1)} \psi _n (x + 1)\\[0.2em]
= \sqrt {\frac{p}{q}(N - n + 1)n} \psi _{n - 1} (x) \\[1em]
\lo  \psi _0 (x) = \sqrt {\frac{{N!p^x q^{N - x} }}{{x!(N - x)!}}} \\[0.5em]
\lo  \psi _n (x) = \sqrt {\frac{{q^n (N - n)!}}{{p^n N!n!}}}  {\prod\limits_{k =
0}^{n - 1} {L^ +  (x,n - 1 - k)} } \psi _0 (x) \\[0.5em]
\lo  L^ +  (x,n - 1)L^ -  (x,n)\psi _n (x) = \frac{p}{q}(N - n + 1)n\psi _n (x) \\[0.5em]
\lo  L^ -  (x,n + 1)L^ +  (x,n)\psi _n (x) = \frac{p}{q}(N - n)(n + 1)\psi _n (x)
\end{eqnarray*}

\bigskip
\noindent {\sl Normalized Meixner functions}
$$ \psi _n (x) = \sqrt {\frac{{\mu ^n (1 - \mu )^\gamma  }}{{n!(\gamma )_n }}} \sqrt
{\frac{{\mu ^x \Gamma (x + \gamma )}}{{\Gamma (x + 1)\Gamma (\gamma )}}} m_n^\gamma  (x)$$
\begin{eqnarray*}
\fl {\rm NM1)}\quad \sqrt {\mu (x + \gamma )(x + 1)} \psi _n (x + 1) + \sqrt {\mu x(x +
\gamma  - 1)} \psi _n (x - 1) - \\[0.5em]
\lo  - \left[ {\mu (x + \gamma ) + x} \right]\psi _n (x) + n(1 - \mu
)\psi _n (x) = 0 \\[1em] 
\fl {\rm NM2)}\quad   - \sqrt {\mu (n + \gamma )(n + 1)} \psi _{n + 1} (x) - \sqrt {\mu n(n + 
\gamma  - 1)} \psi _{n - 1} (x) + \\[0.5em]
\lo  + \left[ {\mu (x + n + \gamma ) + n - x} \right]\psi _n (x) =0 \\[1em] 
\fl {\rm NM3)}\quad  L^ +  (x,n)\psi _n (x) =  - \left[ {\mu (x + n + \gamma )} \right]\psi _n
(x) + \sqrt {\mu x(x + \gamma  - 1)} \psi _n (x - 1) = \\[0.5em]
\lo  = - \sqrt {\mu (n + \gamma )(n + 1)}
\psi _{n + 1} (x) \\[1em]
\fl {\rm NM4})\quad  L^ -  (x,n)\psi _n (x) =  - \left[ {\mu (x + n + \gamma )} \right]\psi _n
(x) + \sqrt {\mu (x + 1)(x + \gamma )} \psi _n (x + 1) = \\[0.5em]
\lo  = - \sqrt {\mu (n + \gamma 
- 1)n} \psi _{n - 1} (x) \\[1em]
\lo  \psi _0 (x) = \sqrt {(1 - \mu )^\gamma  } \sqrt {\frac{{\mu ^x \Gamma (x + \gamma
)}}{{\Gamma (x + 1)\Gamma (\gamma )}}} \\[0.5em]
\lo  \psi _n (x) = \frac{{( - 1)^n }}{{\sqrt {\mu ^n (\gamma )_n n!} }}
{\prod\limits_{k = 0}^{n - 1} {L^ +  (x,n - 1 - k)} } \psi _0 (x) \\[1em]
\lo  L^ +  (x,n - 1)L^ -  (x,n)\psi _n (x) = \mu (n + \gamma  - 1)n\psi _n (x) \\[1em]
\lo  L^ -  (x,n + 1)L^ +  (x,n)\psi _n (x) = \mu (n + \gamma )(n + 1)\psi _n (x)
\end{eqnarray*}

\bigskip
\noindent {\sl Normalized Charlier functions}
$$\quad \psi _n (x) = \sqrt {\frac{{\mu ^n }}{{n!}}} \sqrt {\frac{{e^{ - \mu } \mu ^x
}}{{x!}}} c_n^{(\mu )} (x)$$
\begin{eqnarray*}
\fl {\rm NC1})\quad \sqrt {\mu (x + 1)} \psi _n (x + 1) + \sqrt {\mu x} \psi _n (x - 1) - (x
+ \mu )\psi _n (x) + n\psi _n (x) = 0 \\[1em]
\fl {\rm NC2})\quad  - \sqrt {\mu (n + 1)} \psi _{n + 1} (x) - \sqrt {\mu n} \psi _{n - 1}
(x) + (n + \mu  - x)\psi _n (x) = 0 \\[1em]
\fl {\rm NC3})\quad  L^ +  (x,n)\psi _n (x) =  - \mu \psi _n (x) + \sqrt {\mu x} \psi _n (x -
1) =  - \sqrt {\mu (n + 1)} \psi _{n + 1} (x) \\[1em]
\fl {\rm NC4})\quad  L^ -  (x,n)\psi _n (x) =  - \mu \psi _n (x) + \sqrt {\mu (x + 1)} \psi _n
(x + 1) =  - \sqrt {\mu n} \psi _{n - 1} (x) \\[.5em]
\lo  \psi _0 (x) = \sqrt {\frac{{e^{ - \mu } \mu ^x }}{{x!}}} \\
\lo  \psi _n (x) = \frac{{( - 1)^n }}{{\sqrt {\mu ^n n!} }} {\prod\limits_{k =
0}^{n - 1} {L^ +  (x,n - 1 - k)} } \psi _0 (x) \\[.5em]
 \lo  L^ +  (x,n - 1)L^ -  (x,n)\psi _n (x) = \mu n\psi _n (x) \\[.5em]
 \lo  L^ -  (x,n + 1)L^ +  (x,n)\psi _n (x) = \mu (n + 1)\psi _n (x)
\end{eqnarray*}

\bigskip
\noindent {\sl Normalized Chebyshev functions}
$$\psi _n (x) = \sqrt {\frac{{(2n + 1)(N - n - 1)}}{{(N + n)!}}} t_n (x)$$
\begin{eqnarray*}
\fl {\rm NT1})\quad (x + 1)(N - x - 1) + \psi _n (x + 1) + x(N - x)\psi _n (x - 1) -  \\[.5em] 
 \lo   - \left[ {(x + 1)(N - x - 1) + x(N - x)} \right]\psi _n (x) + n(n + 1)\psi _n (x) = 0
\\[1em]
\fl {\rm NT2})\quad \frac{{n + 1}}{2}\sqrt {\frac{{(2n + 1)(N^2  - n^2  - 2n - 1)}}{{2n +
3}}} \psi _{n + 1} (x) + \frac{n}{2}\sqrt {\frac{{(2n + 1)(N^2  - n^2 )}}{{2n - 1}}} \psi _{n -
1} (x)  \\[.5em]
\lo  +\  (2n + 1)\left( {\frac{{N - 1}}{2} - x} \right)\psi _n (x) = 0 \\[.5em]
\fl {\rm NT3})\quad   L^ +  (x,n)\psi _n (x) =  - \left[ {\frac{1}{2}(n + 1)(N - 2x - n - 1) + x(N -
x)}
\right]\psi _n (x) \\[0.2em]
+ x(N - x)\psi _n (x - 1) = \frac{{n + 1}}{2}\sqrt {\frac{{(2n +
1)(N^2  - n^2  - 2n - 1)}}{{2n + 3}}}
\psi _{n + 1} (x) \\[.5em]
\fl {\rm NT4})\quad  L^ -  (x,n)\psi _n (x) = \Bigg[ \frac{1}{2}(n + 1)(N -2x - n - 1) + n(n + 1) \\[0.2em]
+ (2n + 1)\left( {x - \frac{{N - 1}}{2}} \right)  - \ {(x + 1)(N - x - 1)}\Bigg]\psi _n (x)  \\[0.2em]
+ (x + 1)(N - x - 1)\psi _n (x + 1)  = \frac{n}{2}\sqrt {\frac{{(2n +
1)(N^2  - n^2 )}}{{2n - 1}}} \psi _{n - 1} (x) \\[.5em]
\lo  \psi _0 (x) = \frac{1}{{\sqrt N }}\\[.5em]
\lo \psi _n (x) = \prod\limits_{k = 0}^{n - 1} {\left\{ {\frac{2}{{k + 1}}\sqrt
{\frac{{2k + 3}}{{(2k + 1)(N^2  - k^2  - 2k - 1)}}} L^ +  (x,n - 1 - k)} \right\}\psi _0
(x)}  \\[.5em]
\lo  L^ +  (x,n - 1)L^ -  (x,n)\psi _n (x) = \frac{{n^2 }}{4}(N + n)(N - n)\psi _n (x) \\[.5em]
\lo  L^ -  (x,n + 1)L^ +  (x,n)\psi _n (x) = \frac{{(n + 1)^2 }}{4}(N + n + 1)(N - n -
1)\psi _n (x)
\end{eqnarray*}

\bigskip
\noindent {\sl Normalized Hahn functions}
\begin{eqnarray*}
\psi _n (x) = \sqrt {\frac{{(2n + \alpha  + \beta  + 1)n!(N - n - 1)!\Gamma (n +
\alpha  + \beta  + 1)}}{{\Gamma (n + \alpha  + 1)\Gamma (n + \beta  + 1)\Gamma (N + n +
\alpha  + \beta  + 1)}}} \\[0.2em]
\qquad \qquad \times \sqrt {\frac{{\Gamma (N + \alpha  - x)\Gamma (x + \beta  +
1)}}{{\Gamma (N - x)\Gamma (x + 1)}}} h_n^{(\alpha ,\beta )} (x)
\end{eqnarray*}
\begin{eqnarray*}
\fl {\rm NHa1})\quad   \sqrt {(N - x - 1)(x + \beta  + 1)(N + \alpha  - x - 1)(x + 1)}
\psi _n (x + 1) + \\[.5em]
\lo  + \ \sqrt {(N - x)(x + \beta )(N + \alpha  - x)x} \psi _n (x - 1)  - \bigg\{ N - x - 1)(x + \beta  + 1) \\[0.2em]
\lo + x(N + \alpha  - x) \bigg\}\psi _n (x) + n(n + \alpha  + \beta  + 1)\psi _n (x) = 0 \\[1em]
\fl {\rm NHa2})\quad   \frac{{\sqrt {(n + 1)(n + \alpha  + 1)(n + \beta  + 1)(n + \alpha 
+ \beta  + 1)(2n + \alpha  + \beta  + 1)}}}{{(2n +\alpha  + \beta  + 2)}} \\[0.2em]
\lo \times \sqrt {\frac{(N + n + \alpha  + \beta  + 1)(N - n - 1)}{2n + \alpha  + \beta  + 3}}\cdot  \psi _{n + 1} (x) \\[0.2em]
\lo + \frac{{\sqrt {n(n + \alpha )(n + \beta )(n + \alpha  + \beta )(2n + \alpha  +
\beta  + 1)(N + n + \alpha  + \beta )(N - n)} }}{{(2n + \alpha  + \beta )\sqrt {2n +
\alpha  +\beta  - 1} }} \\[0.2em]
\lo \times \psi _{n - 1} (x) + (2n + \alpha  + \beta  + 1)\Bigg\{ \frac{{2N + \alpha  - \beta  - 2}}{4} \\[0.2em]
\lo + \frac{{(\beta ^2  - \alpha ^2 )(2N + \alpha  + \beta )}}{{4(2n + \alpha  + \beta )(2n +
\alpha  + \beta  + 2)}} - x \Bigg\}\psi _n (x) = 0 \\[1em]
\fl {\rm NHa3)}\quad   L^ +  (x,n)\psi _n (x) = \sqrt {x(N + \alpha  - x)(\beta  + x)(N -
x)}\psi _n (x - 1) - \Big[ \frac{{n + \alpha  + \beta  + 1}}{{2n + \alpha  + \beta  +
2}} \\[0.2em]
\lo \times \bigg\{(\beta  + 1)(N - 1) - (\alpha  + \beta  + 2 + 2n)x + (N - n - \beta  - 2)n
\bigg\} \\[0.2em]
\lo + x(N +\alpha - x) \Big]\psi _n (x) = \frac{{\sqrt {(n + 1)(n + \alpha  + 1)(n + \beta  + 1)(n + \alpha  + \beta  +
1)} }}{{(2n + \alpha 
+ \beta + 2)\sqrt {2n + \alpha  + \beta  + 3} }} \\[0.2em]
\lo \times \sqrt {(2n +\alpha  + \beta  + 1)(N + n + \alpha  + \beta  + 1)(N - n - 1)}\; \cdot \psi _{n + 1} (x) \\[1em] 
\fl {\rm NHa4})\quad   L^ -  (x,n)\psi _n (x) = \sqrt {(x + 1)(N + \alpha  - x - 1)(x +
\beta + 1)(N - x - 1)} \psi _n (x + 1)   \\[1em] 
\lo  + \Bigg[ \frac{{n + \alpha  + \beta  + 1}}{{2n + \alpha  + \beta  + 2}}
\Big\{(\beta  + 1)(N - 1) - (\alpha  + \beta  + 2 + 2n) \;x \\[0.2em]
\lo + (N - n - \beta  - 2)n \Big\} + n(n + \alpha  + \beta  + 1)  + (2n + \alpha  + \beta  + 1) \\[0.2em]
\lo \times \Big( {x - \frac{{2N + \alpha  - \beta -2}}{4} -\frac{{(\beta ^2  - \alpha ^2 )(\alpha  + \beta  + 2N)}}{{4(2n + \alpha  + \beta
)(2n +\alpha  + \beta  + 2)}}} \Big) \\[0.2em]
\lo - (N - x - 1)(x + \beta  + 1) \Bigg]\psi _n (x)  = \frac{{\sqrt {n(n + \alpha )(n + \beta )(n + \alpha  + \beta )} }}{{(2n + \alpha  + \beta )\sqrt {2n +
\alpha  +\beta  - 1} }} \\[0.2em]
\lo \times \sqrt	{(2n + \alpha 
+\beta  + 1)(N + n + \alpha  + \beta )(N - n)} \;\cdot \psi _{n - 1} (x) \\[1em]
\fl \qquad \qquad \psi _0 (x) = \sqrt {\frac{{(\alpha  + \beta  + 1)(N - 1)!\Gamma (\alpha  + \beta 
+1)}}{{\Gamma (\alpha  + 1)\Gamma (\beta  + 1)\Gamma (N + \alpha  + \beta  + 1)}}}
\sqrt{\frac{{\Gamma (N + \alpha  - x)\Gamma (x + \beta  + 1)}}{{\Gamma (N - x)\Gamma (x +
1)}}}\\[1em]
\fl \qquad \qquad \psi _n (x) = \prod\limits_{k = 0}^{n - 1} \Bigg\{ \frac{{(2k + \alpha  + \beta  +
2) }}{{\sqrt {(k + 1)(k + \alpha  + 1)(k + \beta  +
1)(k +\alpha  + \beta  + 1)}}} \\[0.2em]
\lo \qquad \qquad \times \sqrt {\frac{ (2k + \alpha  + \beta  + 3)}{(2k + \alpha  + \beta  + 1)(N + k + \alpha  + \beta  + 1)(N - k - 1)}}\\[0.2em]
\lo \qquad \qquad \times L^ +  (x,n - 1 - k) \Bigg\} \cdot \psi _0  \\[1em]
\fl \qquad \qquad L^ +  (x,n - 1)L^ -  (x,n)\psi _n (x) \\[0.2em]
\lo \qquad \qquad = \frac{{n(n + \alpha )(n + \beta )(n +\alpha 
+ \beta )(N + n + \alpha  + \beta )(N - n)}}{{(2n + \alpha  + \beta )^2 }}\psi _n(x)
\\[1em]
\fl \qquad \qquad L^ -  (x,n + 1)L^ +  (x,n)\psi _n (x) = \frac{{(n + 1)(n + \alpha  + 1)(n + \beta +
1)}}{{(2n + \alpha 
+\beta  + 2)^2 }} \\[0.2em]
\lo \qquad \qquad \times (n + \alpha  + \beta  + 1)(N + n + \alpha  + \beta  + 1)(N - n - 1)\psi _n (x)
\end{eqnarray*}


\ \newline

\end{document}